\documentclass[conference,a4paper]{IEEEtran}
\IEEEoverridecommandlockouts
\usepackage{graphicx,amsmath,epsfig,times}
\usepackage{psfrag,cite}
\usepackage{amstext,amssymb,latexsym}
\usepackage{setspace}

\DeclareMathOperator*{\argmin}{arg\,min}
\begin{document}

\sloppy

\title{Generalizing the Sampling Property of the Q-function for Error Rate Analysis of Cooperative Communication in Fading Channels}

\author{\IEEEauthorblockN{Tu\u{g}can~Akta\c{s}}
\IEEEauthorblockA{Dept. of Electrical and Electronics Eng.\\
Middle East Technical University\\
Ankara, Turkey\\
Email: taktas@metu.edu.tr}
\and
\IEEEauthorblockN{Ali~\"{O}zg\"{u}r~Y\i lmaz}
\IEEEauthorblockA{Dept. of Electrical and Electronics Eng.\\
Middle East Technical University\\
Ankara, Turkey\\
Email: aoyilmaz@metu.edu.tr}
\and
\IEEEauthorblockN{Emre~Akta\c{s}}
\IEEEauthorblockA{Dept. of Electrical and Electronics Eng.\\
Hacettepe University\\
Ankara, Turkey\\
Email: aktas@ee.hacettepe.edu.tr}
}



\maketitle


\begin{abstract}
THIS PAPER IS ELIGIBLE FOR THE STUDENT PAPER AWARD.

This paper extends some approximation methods that are used to identify closed form Bit Error Rate (BER) expressions which are frequently utilized in investigation and comparison of performance for wireless communication systems in the literature. By using this group of approximation methods, some expectation integrals, which are complicated to analyze and have high computational complexity to evaluate through Monte Carlo simulations, are computed. For these integrals, by using the sampling property of the integrand functions of one or more arguments, reliable BER expressions revealing the diversity and coding gains are derived. Although the methods we present are valid for a larger class of integration problems, in this work we show the step by step derivation of the BER expressions for a canonical cooperative communication scenario in addition to a network coded system starting from basic building blocks. The derived expressions agree with the simulation results for a very wide range of signal-to-noise ratio (SNR) values.
\end{abstract}

\section{Introduction}
There has been a great effort for mitigating the detrimental effect of fading in wireless communication systems for a long time. Over the past few years, cooperative communication techniques, which rely on the spatial diversity thanks to the overhearing nodes within the network, have been in the center of these efforts~\cite{Erkip03}. In achieving this cooperation, various methods making use of relay nodes have been proposed~\cite{Laneman04, Laneman06}. The simplest method with demodulate-and-forward (DMF) type relays has an inherent error propagation problem which must be handled at the destination node. Fortunately, when the destination uses the maximum-likelihood (ML) detection~\cite{Laneman06} or cooperative maximal ratio combining (C-MRC)~\cite{Wang07} technique, this problem is overcome and the maximum diversity order is achieved in the system. The canonical cooperative communication scenario, where a single relay node assists the transmission from the source node to the destination, is investigated for the destination with the C-MRC operation in~\cite{Wang07}. The main advantages of C-MRC include its reduced computational complexity with respect to the optimal ML method for increasing number of relays  with practically no loss of BER performance and simpler analysis. 

For the end-to-end BER analysis of C-MRC, in~\cite{Wang07}, the diversity order is considered, which is obtained following a series of detailed upper bounding operations. The closed form BER expressions including the coding gain terms in addition to the diversity order terms would be of importance for this canonical scenario and the one with multiple relays.

Recently in~\cite{Jang2011, Jang2011TCOM}, an approximation method applicable to expectation integrals of a single random variable is proposed for the analysis of fading channel BER performance of cooperative systems. The crucial observation of these papers is that the Q-function appearing in the instantaneous BER function can be well approximated with a Dirac delta well especially in the high SNR region. However, the upper bound utilized on Q-function to obtain the impulse location leads to degradation in the approximation results. More importantly, the analysis lacks rigorous handling of the properties of the sampling function at points other than the critical point. Moreover, the analysis in~\cite{Jang2011, Jang2011TCOM} takes directly the Q-function as its sampling function without any consideration on the SNR value for which the approximation is made. In~\cite{Jang2012}, the SNR regions are identified using the relative error of each possible selection of the sampling function, however a sound analysis of the sampling property is still not provided. 

In this work, we have four goals: (i) A more rigorous method to show the applicability of the sampling property, (ii) Characterization of the SNR regions for sampling property which yields improved approximate expressions, (iii) A more accurate identification of the impulse function parameters, (iv) Generalization of the idea of sampling to integrals involving more than one variables (here for the canonical cooperative scenario, $2$ variables) to obtain closed form BER expressions approximating these integrals. Finally, as an application of sampling property, we sketch the method to obtain closed form expressions that well-approximate the simulated performance of a sample network coded scenario from~\cite{Aktas2013}.

\section{Canonical Cooperative Communication Model and Instantaneous End-to-End BER Expression}
\label{sec:model}
\begin{figure}[htbp]
   \centering
   \includegraphics[width=0.38\textwidth]{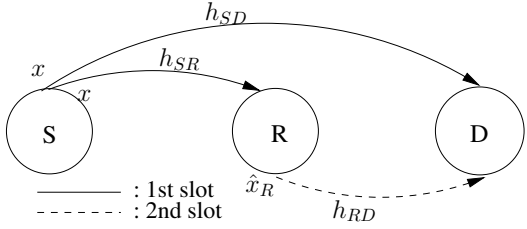}
   \caption{Canonical cooperative communication model}
   \label{fig:model}
\end{figure}

The canonical relay network includes the link from the source node $S$ to the relay node $R$ called $S-R$, the one from $R$ to destination $D$ named $R-D$ in addition to the direct link from $S$ to $D$ shown $S-D$~\cite{Wang07}. The complex channel fading coefficients related to these links are shown in Fig.~\ref{fig:model}. In the first time slot only $S$ transmits the data symbol $x$ and in the second slot only $R$ transmits its own possibly erroneous detection result $\hat{x}_R$.
For the sake of simplicity in analysis, as in~\cite{Wang07}, we assume $x\in \{ +1,-1\}$. In both of the transmissions the average transmit power is taken as $P$. The independent channel fading coefficients $h_{SR}$, $h_{RD}$, and $h_{SD}$ are assumed to follow zero mean circularly symmetric complex Gaussian probability distributions such that $h_{ij}\sim CN(0,\sigma_{ij}^2)$, where $(ij) \in \{SR, RD, SD\}$. Then the received signals at $R$ and $D$ following the first slot are written respectively as
\begin{footnotesize} 
\begin{equation}
\label{eqn:obs}
y_{SR}=h_{SR} x+n_{SR}\ \text{ and } \  
y_{SD}=h_{SD} x+n_{SD},
\end{equation}
\end{footnotesize}%
where $n_{SR}$ and $n_{SD}$ denote the independent white complex Gaussian noise terms at $R$ and $D$ with identical distribution $CN(0,N_0)$. We define $\bar{\gamma}=\frac{P}{N_0}$, then the instantaneous SNR values for $S$-$R$ and $S$-$D$ links are $\gamma_{SR}=\bar{\gamma} \vert h_{SR}\vert ^2$ and $\gamma_{SD}=\bar{\gamma} \vert h_{SD}\vert ^2$. These instantaneous values are exponentially distributed with respective expectations $\bar{\gamma} \sigma_{SR}^2$ and $\bar{\gamma} \sigma_{SD}^2$. In accordance with the DMF method, prior to the transmission in the second slot $R$ should detect $x$ using the optimal (ML) rule as follows.
\begin{footnotesize}
\begin{align}
\label{eqn:detR}
\hat{x}_R=\argmin_{x\in \left\lbrace +1,-1 \right\rbrace } \vert y_{SR} - h_{SR}x\vert ^2,
\end{align}
\end{footnotesize}%
where we assume perfect knowledge of $h_{SR}$ at $R$.

The observation of $D$ after the second time slot is then $y_{RD}=h_{RD} \hat{x}_R+n_{RD}$, where $n_{RD}\sim CN(0,N_0)$ and the corresponding instantaneous SNR $\gamma_{RD}=\bar{\gamma} \vert h_{RD}\vert ^2$ is also exponentially distributed. The aim of $D$ is to correctly detect the data symbol $x$ of node $S$ by making use of all instantaneous channel gains. If we follow the C-MRC technique to linearly combine two observations at $D$ as suggested in~\cite{Wang07}, we obtain the following detection result.
\begin{footnotesize}
\begin{equation}
\label{eqn:detD}
\hat{x}_D=\argmin_{x\in \{ +1,-1\} } \vert w_1 y_{SD} + w_2 y_{RD} - (w_1 h_{SD} + w_2 h_{RD})x\vert ^2,
\end{equation}
\end{footnotesize}%
where $w_1$ is the weight coefficient corresponding to the $S-D$ link and selected to be equal to $h_{SD}^\ast$ just as in the well-known MRC case with no relaying operation. On the other hand, for the observation $y_{RD}$, which corresponds to the relayed communication over the links $S-R$ and $R-D$, the coefficient $w_2$ should be re-defined to reflect the possible error propagation on these two hops. Authors, in~\cite{Wang07}, propose a single equivalent channel for representing these hops and assign the equivalent instantaneous SNR of this channel as
\begin{footnotesize}
\begin{equation}
\label{eqn:gamma_eq}
\gamma_{eq}\triangleq \left\lbrace Q^{-1}\left( \left[1-P_{SR}^b\right] P_{RD}^b + \left[1-P_{RD}^b\right] P_{SR}^b\right)\right\rbrace^2 /2, 
\end{equation}
\end{footnotesize}%
where $Q(x)=\int_{x}^{\infty}\frac{1}{\sqrt{2\pi}}\exp(-z^2/2)\text{d}z$, $P_{SR}^b=Q(\sqrt{2\gamma_{SR}})$ and $P_{RD}^b=Q(\sqrt{2\gamma_{RD}})$. If we further define $w_2 =\frac{\gamma_{eq}}{\gamma_{RD}}h_{RD}^\ast$, the instantaneous end-to-end BER expression is found as~\cite{Wang07}
\begin{footnotesize}
\begin{align}
\label{eqn:P_b}
P^b =& \left( 1 - Q(\sqrt{2\gamma_{SR}})\right)Q\left[ \frac{\sqrt{2}(\gamma_{SD}+\gamma_{eq})}{\sqrt{\gamma_{SD}+\gamma_{eq}^2/\gamma_{RD}}}\right]\nonumber\\
+&Q(\sqrt{2\gamma_{SR}})Q\left[ \frac{\sqrt{2}(\gamma_{SD}-\gamma_{eq})}{\sqrt{\gamma_{SD}+\gamma_{eq}^2/\gamma_{RD}}}\right].
\end{align}
\end{footnotesize}%
Clearly, it is analytically hard to evaluate a triple integral for the instantaneous BER given in (\ref{eqn:P_b}) over related distributions of the random variables $\gamma_{SR}$, $\gamma_{SD}$, and $\gamma_{RD}$ (and $\gamma_{eq}$ which is a complicated function of $\gamma_{SR}$ and $\gamma_{RD}$) to reach the average BER value. In~\cite{Wang07}, only the diversity order for the BER-SNR curve is identified following a set of upper bounding techniques and the result is $2$ since two transmissions are made on independent paths in the network and $D$ accounts for possible relaying errors. In Section \ref{sec:Qfunc_relay}, we present novel closed form expressions for the average BER of this system such that the coding and the diversity gains are identified separately.

\section{Sampling Property of the Q-Function for Generalized Expressions}
\label{sec:Qfunc}
This section starts with the basic result of~\cite{Jang2011TCOM} with a slight improvement on obtaining the critical point of the related sampling function. We also exemplify the problem with this technique for the low SNR region and propose a method to identify threshold value of SNR over which the technique is applicable. We further generalize the sampling property to obtain good approximations for lower SNR values. Finally, we generalize this sampling property to integrals involving more than one variables.
\subsection{Basic Problem and its Solution}
\label{sec:Qfunc_single}
Assume that the following expectation integral of an instantaneous probability of error function $Q(\sqrt{X})$ is to be evaluated for a random variable $X$ with pdf $f_X(x)$~\cite{Jang2011TCOM}.
\begin{footnotesize} 
\begin{equation}
\label{eqn:Qsqrtx}
I_0 = E_X\left\lbrace Q(\sqrt{X}) \right\rbrace = \int_{0}^{\infty}Q(\sqrt{x})f_X(x)\text{d}x.
\end{equation}
\end{footnotesize}%
After the change of variables operation $x\rightarrow t^N$ , we have the integrand $Q(\sqrt{t^N})Nt^{N-1}f_X(t^N)$. Here we define the following constituent functions of the integrand.
\begin{equation}
\label{eqn:prod_defn}
q(t)\triangleq Q(\sqrt{t^N}), \;c(t) \triangleq Nt^{N-1},\; f(t)\triangleq f_X(t^N).
\end{equation}
In~\cite{Jang2011TCOM}, $h(t^N)\triangleq q(t)c(t)$ is defined and claimed to be a unimodal function of $t$ with a critical point satisfying $t^N_{\ast}=2$. However, this analysis does not show that $h(t^N)$ takes value of zero for any other value and on the contrary, when any other finite $t^N$ value is inserted in Eqn. (52) of~\cite{Jang2011TCOM} it is easy to show that  $h(t^N)$ assumes infinity. Moreover, the critical point $t^N_{\ast}=2$ is obtained after applying an upper bound on the Q-function. Let us start with the analysis of the integrand $I(t)\triangleq q(t) c(t) f(t)$ for three distinct regions of $t$.
\begin{small}
\begin{equation}
\label{eqn:Qsqrtx_3region}
I_0 = \int_{0}^{1^-}I(t)\text{d}t + \int_{1^-}^{1^+}I(t)\text{d}t + \int_{1^+}^{\infty}I(t)\text{d}t.
\end{equation}
\end{small}%
It should be emphasized that the value $I_0$ in (\ref{eqn:Qsqrtx_3region}) is independent of $N$ and hence we may investigate the behaviour of $I(t)$ asymptotically (for ${N\rightarrow\infty}$). Moreover, in this work, we take $f(t)=\frac{1}{SNR}e^{-\frac{t^N}{SNR}}$ (Rayleigh fading assumption) although the following steps can be generalized to other pdfs. Initially, for the $0<t<1$ region, we have 
\begin{small}
\begin{align}
\label{eqn:t_less_1}
\lim_{N\rightarrow\infty}I(t) \bigg\vert_{0<t<1}&= \left(\lim_{N\rightarrow\infty}q(t)\right) \left(\lim_{N\rightarrow\infty}c(t)\right) \left(\lim_{N\rightarrow\infty}f(t)\right)\nonumber\\
&= \left(\frac{1}{2}\right) \left(\lim_{N\rightarrow\infty}\frac{1}{\ln{t} (-t^{1-N})}\right) \left(\frac{1}{SNR}\right) \;\nonumber\\
&= 0,
\end{align}
\end{small}%
where the L'H\^{o}pital rule is applied for $\lim_{N\rightarrow\infty}c(t)$ term. As a result, the first integral in (\ref{eqn:Qsqrtx_3region}) evaluates to $0$ asymptotically. Next, for the $t>1$ region, we have 
\begin{small}
\begin{align}
\label{eqn:t_larger_1}
\lim_{N\rightarrow\infty}I(t) \bigg\vert_{t>1} &= \left(\lim_{N\rightarrow\infty}q(t)\right) \left(\lim_{N\rightarrow\infty}c(t)f(t)\right)\nonumber\\
&= \left(0\right) \left(\lim_{N\rightarrow\infty}\frac{N t^{N-1}}{\exp\left( \frac{t^N}{SNR}\right)}\right) \left(\frac{1}{SNR}\right)\nonumber\\
&= \left(0\right) \left(\lim_{N\rightarrow\infty}\frac{t^{N-1} + N (\ln{t}) t^{N-1}}{\frac{\ln{t}}{SNR}t^N\exp\left( \frac{t^N}{SNR}\right)}\right)\nonumber\\
&= \left(0\right) \left(\lim_{N\rightarrow\infty}\frac{\ln{t}}{\left(\frac{\ln{t}}{SNR}\right)^2 t^N\exp\left( \frac{t^N}{SNR}\right)}\right) \nonumber\\
&= 0
\end{align}
\end{small}%
following application of the L'H\^{o}pital rule twice. Hence, in the asymptotic sense, the third integral in (\ref{eqn:Qsqrtx_3region}) does not contribute to the result as well. Therefore, we obtain
\begin{small}
\begin{equation}
\label{eqn:Qsqrtx_1region}
I_0 = \int_{1^-}^{1^+}\lim_{N\rightarrow\infty} I(t)\text{d}t,
\end{equation}
\end{small}%
which shows us that the integrand $I(t)$ may be well-approximated by a Dirac delta at $t=1$ for ${N\rightarrow\infty}$. Moreover, it can be shown that this Dirac delta approximation also holds for the functions $h(t^N)\triangleq q(t)c(t)$ and $g(t^N)\triangleq f(t)c(t)$. Herein we are going to assess the limiting SNR value above which $h(t^N)$ can be safely approximated by a Dirac delta. In order to obtain this threshold, we are going to compare convergence rates of $h(t^N)$ and $g(t^N)$ in Section~\ref{sec:comp_conv}.
\subsection{Rates of Convergence for Constituent Functions}
\label{sec:comp_conv}
If we remember (\ref{eqn:Qsqrtx_3region}), two regions that asymptotically yield no contribution to the integral should be examined for convergence rate comparison. For the $0<t<1$ region, both $q(t)$ and $f(t)$ converge to nonzero constants in the limit. On the other hand, the function $c(t)=Nt^{N-1}$ converges to $0$, which means that for large $N$, $h(t^N)$ and $g(t^N)$ converge to $0$ with the same rate. Hence we will concentrate on the other region: $t>1$. In this region we are going to compare $h(t^N)$ and $g(t^N)$ starting with the Chernoff upper bound on the Q-function.
\begin{small}
\begin{equation}
\label{eqn:q_chernoff}
q(t)\bigg\vert_{t^N=x} = Q(\sqrt{x}) \leq \frac{1}{2} \exp\left( -\frac{x}{2}\right).
\end{equation}
\end{small}%
In addition, for $t>1$ and $SNR \geq 2$ it is easily shown that 
\begin{small}
\begin{equation}
\label{eqn:exp_ineq}
f(t)\bigg\vert_{t^N=x} = \frac{\exp\left(\frac{-x}{SNR}\right)}{SNR} \geq \frac{1}{2} \exp\left( -\frac{x}{2}\right).
\end{equation} 
\end{small}%
Combining (\ref{eqn:q_chernoff}) and (\ref{eqn:exp_ineq}) for $t>1$ and $SNR \geq 2$ (in dB scale roughly for values larger than $3$ dB) we get
\begin{footnotesize}
\begin{equation}
\label{eqn:q_exp_comp}
q(t)\bigg\vert_{t^N=x} \leq f(t)\bigg\vert_{t^N=x}.
\end{equation}
\end{footnotesize}%
Using (\ref{eqn:q_exp_comp}) we reach the result that for $SNR>2$, $h(t^N)$ (including the Q-function) is better represented by a Dirac delta with respect to $g(t^N)$ (including the exponential pdf). Also, for $x>1$, one can show that 
\begin{footnotesize}
\begin{equation}
\label{eqn:q_exp_comp2}
q(t)\bigg\vert_{t^N=x} = Q(\sqrt{x}) \geq \frac{1}{\frac{1}{3}} \exp\left( -\frac{x}{\frac{1}{3}}\right) \geq \frac{\exp\left(\frac{-x}{SNR}\right)}{SNR}=f(t)\bigg\vert_{t^N=x}.
\end{equation}
\end{footnotesize}%
taking $SNR<1/3$ (roughly less than $-5$ dB). Consequently, for lower SNR values, $g(t^N)$ fits better to sampling function definition. Firstly, the position of the Dirac delta that approximates $g(t^N)$ can be obtained by finding the critical point $t_g ^N$. We equate the first derivative of $g(t^N)$ with respect to $t$ to $0$:
\begin{small}
\begin{equation}
\frac{\text{d}}{\text{d}t}g(t^N)\bigg\vert_{t^N=t_g ^N}=0,
\end{equation}
\end{small}%
whose solution is 
\begin{small}
\begin{equation}
\label{eqn:exp_critical}
t_g ^N = \frac{N-1}{N}SNR.
\end{equation}
\end{small}%
Eqn. (\ref{eqn:exp_critical}) gives us the asymptotic critical point $t_\ast ^N = \lim_{N\rightarrow\infty} t_g ^N=SNR$. Secondly, the weight of the corresponding Dirac delta is found as $1$ due to the normalization property of the pdf. Hence for $SNR<1/3$, we may use the approximation $I_0\approx \int_{0}^{\infty}Q(\sqrt{x})\delta(x-SNR)\text{d}x = Q\left( \sqrt{SNR}\right)$.

\begin{figure}[htbp]
   \centering
   \includegraphics[width=0.48\textwidth]{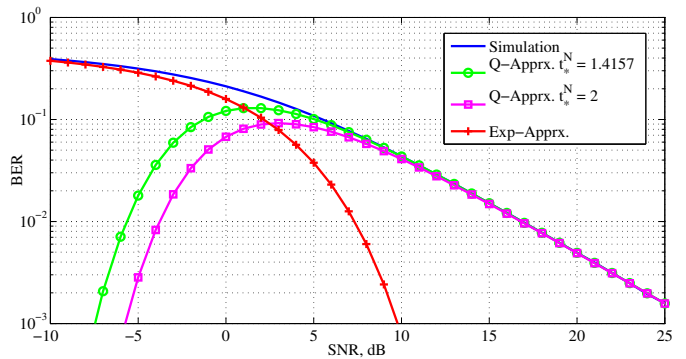}
   \caption{Approximating the integral $I_0$ using various methods}
   \label{fig:q_exp_comp}
\end{figure}

For $SNR>2$, we write $I_0\approx \int_{0}^{\infty}c\delta(x-t_\ast ^N)f_X(x)\text{d}x = \frac{c}{SNR}\exp\left( -\frac{t_\ast ^N}{SNR}\right)$, where the impulse weight is found using the alternative definition of Q-function as $c=\int_{0}^{\infty}h(t^N)\text{d}t=\int_{0}^{\infty}Q(\sqrt{x})\text{d}x=\frac{1}{2}$. We propose a simple alternative to the method in~\cite{Jang2011TCOM} to evaluate the location of the impulse approximating $h(t^N)$. For a sufficiently large value of $N$, we pose finding the critical point as an unconstrained optimization problem and employ numerical search to find the solution. In a narrow neighbourhood of $t=1$, we search for the critical point of $h(t^N)$ and we find $t_\ast ^N = 1.4157$ just after $4$ iteration steps, whereas in~\cite{Jang2011TCOM} $t_\ast ^N = 2$ was considered. The approximation methods as well as the simulation result are given in Fig.~\ref{fig:q_exp_comp}.
According to Fig.~\ref{fig:q_exp_comp}, the methods approximating $h(t^N)$ as a Dirac delta (one proposed in~\cite{Jang2011TCOM} with square markers, and the one we propose with circular markers) are quite consistent for high SNR values. However the method using unconstrained optimization for searching impulse location is the better one with close approximation for $SNR>3dB$ as detailed in equation (\ref{eqn:q_exp_comp}). For low SNR values, on the other hand, only the method selecting $g(t^N)$ as the sampling function (plus shaped markers) is in a sensible proximity to the simulation result. This shows us that for a close approximation of the integral $I_0$ over the whole SNR region, we need to use a piecewise function as a result of (\ref{eqn:q_exp_comp}) and (\ref{eqn:q_exp_comp2}). In the following sections, with more than one variables case, we are going to use only the sampling property for the Q-function since we are mostly interested in high SNR performance.

\subsection{Two-Variable Sampling Property}
\label{sec:Qfunc_double}
\begin{figure}[htbp]
   \centering
   \includegraphics[width=0.38\textwidth]{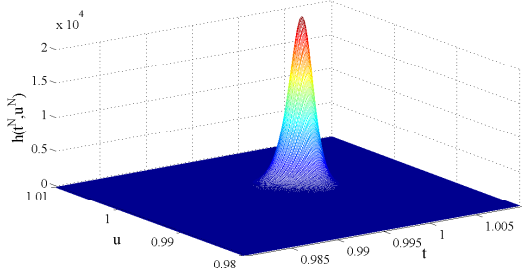}
   \caption{Function $h(t^N,u^N)$ for $N=1000$}
   \label{fig:Qsqrtx_y}
\end{figure}

In this section we are going to base our discussion on the following integral involving two variables in the integrand.
\begin{equation}
\label{eqn:Qsqrtx_y}
I_1 = \int_{0}^{\infty}\int_{0}^{\infty}Q(\sqrt{a_1x +a_2y})f_{X}(x)f_{Y}(y)\text{d}x\text{d}y, 
\end{equation}
where $a_1$ and $a_2$ are positive constants. Similar to the single dimension analysis we define $h(t^N,u^N)\triangleq Q(\sqrt{a_1t^N +a_2u^N})N^2t^{N-1}u^{N-1}$ based on the Q-function. Here, it is easy to generalize the asymptotic analysis for $h(t^N,u^N)$ with $N\rightarrow\infty$ to show that it is well-approximated by a two-dimensional Dirac delta at $(t,u)=(1,1)$. This is further exemplified in Fig.~\ref{fig:Qsqrtx_y} for $N=1000$ and $a_1=a_2=2$. Such a higher dimensional generalization for sampling property is not made in~\cite{Jang2011,Jang2011TCOM,Jang2012}, but whenever a two-dimensional integrand is encountered, the double integral is approximated by two single variable integrals yielding coding gain errors in the final expressions.
Through unconstrained optimization solution, the critical point of $h(t^N,u^N)$ is computed as $(t^N_{\ast},u^N_{\ast}) =(0.8197,0.8197)$ and the weight of the Dirac delta is analytically found as $c=\int_{0}^{\infty}\int_{0}^{\infty}Q(\sqrt{a_1x +a_2y})\text{d}x\text{d}y=\frac{3}{4a_1a_2}$. As an example for $a_1=a_2=2$, the approximation for $I_1$ is
\begin{footnotesize}
\begin{equation}
\label{eqn:Qsqrtx_y2}
I_1 \approx \frac{3}{16SNR^2}\exp\left(-\frac{2(0.8197)}{SNR}\right).
\end{equation}
\end{footnotesize}%
The result in (\ref{eqn:Qsqrtx_y2}) is in accordance with the average BER analysis result of the MRC technique applied on two parallel branches~\cite{Goldsmith} and is also very close to the simulation result for mid- to high SNR values as given in Fig.~\ref{fig:Qcombined}.

\subsection{Sampling in Single Dimension for Functions of Two Variables}
\label{sec:Qfunc_min}
Unfortunately, not every integrand function can be simply approximated with a two dimensional Dirac delta as in Section~\ref{sec:Qfunc_double}. As an example, the instantaneous BER function $Q(\sqrt{2\min\{x,y\}})$ can be shown to diverge for some points other than $(t,u)=(1,1)$ following the suitable change of variables operation. However, it is still possible to analyze this function using the fact that $Q(\sqrt{2\min\{x,y\}})\leq Q(\sqrt{2x}) + Q(\sqrt{2y})$ and the sampling property for single variable functions given in Section~\ref{sec:Qfunc_single}. Then we can reach the following approximation for this expectation integral, which is shown to perfectly fit the simulation result in Fig.~\ref{fig:Qcombined}.
\begin{footnotesize}
\begin{align}
\label{eqn:Qsqrtminx_y}
I_2 =& \int_{0}^{\infty}\int_{0}^{\infty}Q(\sqrt{2\min\{x,y\}})f_{X}(x)f_{Y}(y)\text{d}x\text{d}y\nonumber\\
\approx& \frac{1}{2SNR}\exp \left( -\frac{0.7079}{SNR}\right).
\end{align}
\end{footnotesize}%

\subsection{BER analysis for the Canonical Cooperative Model}
\label{sec:Qfunc_relay}
\begin{figure}[htbp]
   \centering
   \includegraphics[width=0.488\textwidth]{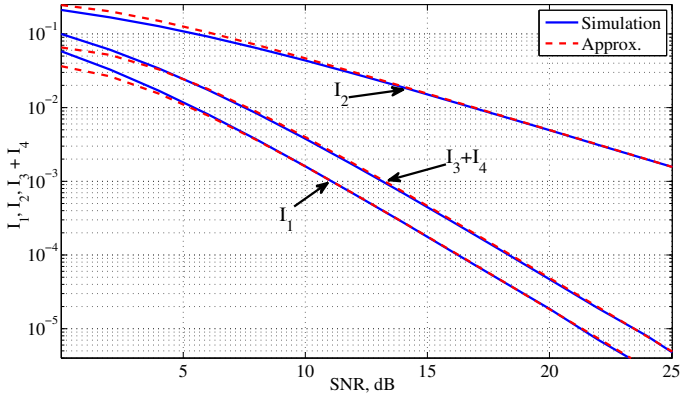}
   \caption{Approximating integrals $I_1$, $I_2$, and $I_3+I_4$}
   \label{fig:Qcombined}
\end{figure}

The end-to-end instantaneous BER function in (\ref{eqn:P_b}) can be written as the sum of two terms: $P^b=P_1+P_2$. Let us start with $P_1 = \left( 1 - Q(\sqrt{2\gamma_{SR}})\right)Q\left[ \frac{\sqrt{2}(\gamma_{SD}+\gamma_{eq})}{\sqrt{\gamma_{SD}+\gamma_{eq}^2/\gamma_{RD}}}\right]$, which is a function of three variables, $\gamma_{SR}$, $\gamma_{RD}$, and $\gamma_{SD}$. Similar to the analysis in Section~\ref{sec:Qfunc_min}, function $P_1$ can not be approximated with a Dirac delta directly, due to the variable $\gamma_{eq}$ defined over the instantaneous SNR values of the two-hop link, $\gamma_{RD}$ and $\gamma_{SD}$. Therefore, we define two terms that are asymptotic in $\gamma_{RD}$ and $\gamma_{SD}$ following the approach in Eqn. (42) of~\cite{Jang2011}
\begin{footnotesize}
\begin{align}
\label{eqn:P_1_lim}
P_1^{\gamma_{RD}} \triangleq & \lim_{\gamma_{RD}\to\infty} P_1 = \left( 1 - Q(\sqrt{2\gamma_{SR}})\right)Q\left[ \frac{\sqrt{2}(\gamma_{SD}+\gamma_{SR})}{\sqrt{\gamma_{SD}}}\right]\nonumber\\
P_1^{\gamma_{SR}} \triangleq & \lim_{\gamma_{SR}\to\infty} P_1 = Q\left[ \sqrt{2(\gamma_{SD}+\gamma_{RD})}\right]
\end{align}
\end{footnotesize}%
to approximate $P_1$ with the sum of these two terms. In this way, $P_1$ is now the sum of two functions both of which have two arguments and are suitable for an approximation with impulse functions. It should be noted that in the approach utilized in Section~\ref{sec:Qfunc_min}, $Q(\sqrt{2x})$ and $Q(\sqrt{2y})$ are also asymptotic terms. Using the result of Section~\ref{sec:Qfunc_double}, approximate expectation of $P_1$ is evaluated to be
\begin{footnotesize}
\begin{align}
\label{eqn:P_1}
I_3 \approx &  \int_{0}^{\infty} \int_{0}^{\infty} P_1^{\gamma_{RD}}f_{\gamma_{SR}}(\gamma_{SR})f_{\gamma_{SD}}(\gamma_{SD})\text{d}\gamma_{SR}\text{d}\gamma_{SD}\nonumber\\
+& \int_{0}^{\infty} \int_{0}^{\infty} P_1^{\gamma_{SR}}f_{\gamma_{RD}}(\gamma_{RD})f_{\gamma_{SD}}(\gamma_{SD})\text{d}\gamma_{RD}\text{d}\gamma_{SD}\nonumber\\
\approx &\frac{1}{16SNR^2}\exp \left( -\frac{1.3049}{SNR}\right)+\frac{3}{16SNR^2}\exp \left( -\frac{2(0.8197)}{SNR}\right).
\end{align}
\end{footnotesize}%
Defining similar asymptotic terms for $P_2$, we reach
\begin{footnotesize}
\begin{align}
\label{eqn:P_2}
I_4 \approx &  \int_{0}^{\infty} \int_{0}^{\infty} P_2^{\gamma_{RD}}f_{\gamma_{SR}}(\gamma_{SR})f_{\gamma_{SD}}(\gamma_{SD})\text{d}\gamma_{SR}\text{d}\gamma_{SD}\nonumber\\
\approx &\frac{1}{4SNR^2}\exp \left( -\frac{1.7564+1.3737}{SNR}\right).
\end{align}
\end{footnotesize}%
Finally, summing the results of (\ref{eqn:P_1}) and (\ref{eqn:P_2}) we obtain and plot the approximate expectation of $P^b$ as $I_3+I_4$ in Fig.~\ref{fig:Qcombined} together with the simulation result. It is seen that the analysis proposed in this work yields an extremely good approximation to the end-to-end average BER of the canonical cooperative communication system by giving the closed form expression as a product of the coding and diversity gain terms.

\section{BER Analysis for a Network Coded System}
\label{sec:network}
\begin{figure}[htbp]
   \centering
   \includegraphics[width=0.47\textwidth]{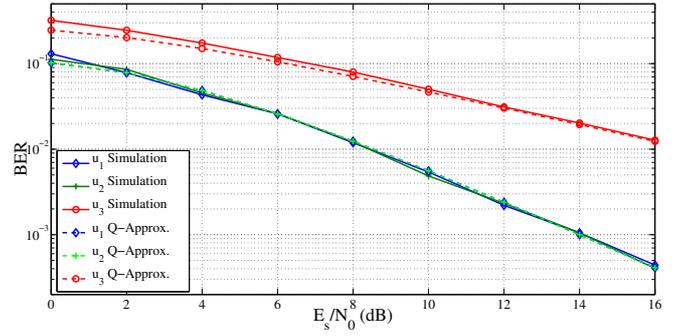}
   \caption{Approximating the average BER for the network coded system}
   \label{fig:equiv_analysis}
\end{figure}

In this section, we are going to sketch the BER analysis for a sample network coded cooperative system (given by Eqn. (1) and Fig.1 of~\cite{Aktas2013}) with single destination node and $3$ source nodes that transmit network coded data in $4$ orthogonal transmission slots. This DMF-type relaying system model is essentially the same with the canonical cooperative model of~\cite{Wang07} with the additional XOR operations at the intermediate (relay) nodes and extra time slots. Utilizing the equivalent channel approach for the network coded data, one may reach the following approximate instantaneous probability of error function for the data bit of node $1$,i.e., $u_1$.
\begin{footnotesize}
\begin{align}
\label{eqn:inst_ber}
 P&(\hat{\hat{u}}_1 \neq u_1) \approx (1-p_{e4}) Q\left( \frac{\sqrt{2}\left( \vert h_{1}  \vert ^2 + \vert h_{eq4}  \vert ^2 \right)}{\sqrt{\vert h_{1} \vert ^2 + \frac {\vert h_{eq4}  \vert ^4}{\vert h_{4}  \vert ^2}}} \right) \nonumber\\
 +& p_{e4} Q\left( \frac{\sqrt{2}\left( \vert h_{1}  \vert ^2 - \vert h_{eq4}  \vert ^2 \right)}{\sqrt{\vert h_{1} \vert ^2 + \frac {\vert h_{eq4}  \vert ^4}{\vert h_{4}  \vert ^2}}} \right) + Q\left(\sqrt{2\left( \vert h_{1}  \vert ^2 + \vert h_{2}  \vert ^2\right)} \right),
\end{align}
\end{footnotesize}%
where $p_{e4}$ is the probability of error for the transmitted network coded data in the fourth time slot, $\vert h_{1} \vert ^2$, $\vert h_{2} \vert ^2$, and $\vert h_{4} \vert ^2$ denote the independent instantaneous SNR values for the transmissions in the corresponding slots from the intermediate nodes to the destination node and follow exponential distributions with mean value equal to $SNR$. Similar to the equivalent channel definition for the canonical cooperation model, $\vert h_{eq4} \vert ^2$ is defined to represent the two-hop transmission that ends in the fourth time slot. We immediately make use of the sampling properties derived in Section~\ref{sec:Qfunc_double} to approximate the integral in (\ref{eqn:inst_ber}) and after a few steps we obtain
\begin{footnotesize}
\begin{align}
\label{eqn:avg_ber}
 E&_{h_1,h_2,h_4,p_{e4}}\left\lbrace  P(\hat{\hat{u}}_1 \neq u_1)\right\rbrace \approx \frac{1}{16SNR^2}\exp \left( -\frac{1.3049}{SNR}\right)\nonumber\\
 +& \frac{3}{8SNR^2}\exp \left( -\frac{2(0.8197)}{SNR}\right)
 + \frac{4}{16SNR^2}\exp \left( -\frac{3.1301}{SNR}\right).
\end{align}
\end{footnotesize}%
One can distinguish the diversity order of $2$ and the coding gain using the closed form approximation (\ref{eqn:avg_ber}). Following the same procedure for other two nodes' data bits, we plot Fig.~\ref{fig:equiv_analysis} in order to visualize the approximation results for the network coded system. Even for low SNR values, the proposed Dirac delta approximation for multi-dimensional integrals is in a good agreement with the simulation results.

\section{Conclusion}
In this work we rigorously investigate the applicability of one dimensional sampling property to some frequently encountered expectation integrals using asymptotic analysis for distinct regions of the argument of the integrand functions. Identification of the threshold SNR value for selection of the most suitable part of this integrand gives us the opportunity to define piecewise approximations that are very close to the simulation results over the whole SNR region. Moreover, we extend our sampling property analysis to higher dimensions and through some basic examples, we build up an approximation method to give closed form expressions for the performance evaluation of a canonical cooperative communication scenario. These extensions are further shown to cover a sample network coded scenario yielding closed form expressions that are very close to the simulation results.

\bibliography{wireless_network_coding_v7}

\end{document}